%% file: main.tex
\pdfoutput=1
\documentclass[conference]{IEEEtran}
\usepackage[keeplastbox]{flushend}
\usepackage[latin1]{inputenc}
\usepackage{times,amsmath}
\usepackage{amsfonts}
\usepackage{pstool}
\usepackage{subfigure}
\usepackage{multirow}
\usepackage{multicol}
\usepackage{enumerate}
\usepackage{graphicx}
\usepackage{mathtools, cuted}
\usepackage{MnSymbol}
\usepackage{stfloats}
\usepackage[table]{xcolor}
\usepackage[square, comma, sort&compress, numbers]{natbib}
\usepackage{nohyperref}
\usepackage{algorithm,algorithmic}
\usepackage{bigints}
\usepackage{amsmath}

\newcounter{tempEquationCounter}
\newcounter{thisEquationNumber}

\makeatletter
\newcommand{\vast}{\bBigg@{4}}
\newcommand{\Vast}{\bBigg@{5}}
\newcommand\numeq[1]%
  {\stackrel{\scriptscriptstyle(\mkern-1.5mu#1\mkern-1.5mu)}{=}}
\makeatother

\graphicspath{ {Figures/ } }

\begin{document}


\title{Countering Active Attacks on RAFT-based IoT Blockchain Networks}

\author{
\IEEEauthorblockN{ Hasan Mujtaba Buttar$^\ast$, Waqas Aman$^\ast$, M. Mahboob Ur Rahman$^\ast$, Qammer H. Abbasi$^\dagger$ }
\IEEEauthorblockA{
$\ast$ Electrical engineering department, Information Technology University, Lahore 54000, Pakistan. \\
$^\dagger$ Department of Electronics and Nano Engineering, University of Glasgow, Glasgow, G12 8QQ, UK. \\
$^\ast$\{hasan.mujtaba, waqas.aman, mahboob.rahman\}@itu.edu.pk} $^\dagger$Qammer.Abbasi@glasgow.ac.uk
}

\maketitle

\input{abstract}

\begin{IEEEkeywords}
Blockchain, RAFT consensus, IoT Blockchain Networks, Jamming, Impersonation, Authentication, Coverage Probability, Security, Uplink, Downlink, Stochastic Geometry, Pathloss.
\end{IEEEkeywords}

\input{sec1}

\input{sec2}

\input{sec3}
\input{conclusion}

\appendices
\input{appendix}

\footnotesize{
\bibliographystyle{IEEEtran}
\bibliography{references}
}

\vfill\break

\end{document}

%% file: abstract.tex
\begin{abstract}

This paper considers an Internet of Thing (IoT) blockchain network consisting of a leader node and various follower nodes which together implement the RAFT consensus protocol to verify a blockchain transaction, as requested by a blockchain client. Further, two kinds of active attacks, i.e., jamming and impersonation, are considered on the IoT blockchain network due to the presence of multiple {\it active} malicious nodes in the close vicinity. When the IoT network is under the jamming attack, we utilize the stochastic geometry tool to derive the closed-form expressions for the coverage probabilities for both uplink and downlink IoT transmissions. On the other hand, when the IoT network is under the impersonation attack, we propose a novel method that enables a receive IoT node to exploit the pathloss of a transmit IoT node as its fingerprint to implement a binary hypothesis test for transmit node identification. To this end, we also provide the closed-form expressions for the probabilities of false alarm, missed detection and miss-classification. Finally, we present detailed simulation results that indicate the following: i) the coverage probability improves as the jammers' locations move away from the IoT network, ii) the three error probabilities decrease as a function of the link quality.
\\
\end{abstract}

%% file: sec1.tex
\section{Introduction}
\label{sec:intro}

 The next generation wireless networks are supposed to have support for processing at the edge, automation and distributed trust, which can be accomplished using blockchain enabled wireless networks \cite{Zhang:IWC:2021}.
  Blockchain  technology has great capability in wireless networks (specifically, Internet of thing (IoT) devices) for  developing trust and consensus procedures without the intervention of central party \cite{nguyen2019blockchain}. Recently, some prominent works are reported in the literature studying blockchain based wireless networks. For the first time, Y.Sun in \cite{sun2019blockchain} study the wirelessly connected blockcahin system where he provides optimal nodes deployment, and a relation of communication and transaction throughput. Further, the work \cite{xu2020raft} studies the security aspects of RAFT based wireless networks with the presence of single jammer and derive the probability of achieving successful blockchain transactions via exploiting probability theory. Next, authors in \cite{Zhang:ICC:2020} maximize the transaction uploading and revenue of miners using Stackelberg game approach in blockchain based wireless networks. Following \cite{xu2020raft}, authors in \cite{zhuz2020blockchain} provide theoretical calculation of required throughput and transmission successful probability to support the system.  Further Hao in his perspective article \cite{Hao:DCN:2020} discusses the potentials of blockchain for resource management in 6G. Particularly, he discusses  device to device, IoT, network slicing applications of 6G for resource management.
  More recently, \cite{zhang2:arXiv:2021} answer the question that how much resources are needed to run a wireless blockchain networks. Particularly, the impact of provision of resources is studied on the performance of blockchain. Dynamic spectrum sharing is studied in  \cite{Liu:IWCMC:2021}, where reinforcement learning is utilized to analyze the resource sharing structure and spectrum sharing process in blockchain system combined with 6G hybrid cloud. Next, the authors in \cite{Xuefei;IOTJ;2021} study the block propagation in blochchain based vehicular ad-hoc networks. They study the dynamic of block propagation and provide a closed-form expression for block propagation time. The work \cite{Hou:ICL:2021} minimize the latency in storing data by intelligent transaction migration policy by exploiting Markov process and deep deterministic policy gradient. The authors in \cite{Li:TVT:2021} focus blochchain based wireless local area network and provide a new medium access control protocol known as block access control. Specifically, its design is studied with, modeling and analyses. Finally, the authors in \cite{Hao:Arxiv:2021} propose Blockchain Enabled Radio Access Network (BE-RAN). They provide a security framework for mutual authentication based on digital signatures/secret keys. They also provide design guidelines for switching, routing and quality of service management.

 The ground basis of blockchain is the consensus techniques that establish trust and update the ledger's states which can be broadly  classified into two classes \cite{nguyen2018survey}. The first class, which is based on pure computation, solve the mathematical puzzle by joining nodes of the blockchain network to prove that they are eligible for mining work, e.g., Proof-of-Work (PoW) \cite{peer} and Proof of Stacks (PoS) \cite{vasin2014blackcoin}. The second class relies on the pure communications between joining nodes. The successful voting of majority nodes through the communication channel leads to the achievement of consensus, i.e. Byzantine Fault Tolerance (BFT) \cite{castro1999practical}, Paxos \cite{lamport2019part}, and RAFT \cite{ongaro2014search}. Private blockchains use second class due to its low complex nature, high throughput and small confirmation delay \cite{dinh2018untangling}.
Low complex consensus techniques might be used in wirelessly connected IoT devices to develop the distributed system with the presence of unreliable nodes. In such network, blockchain synchronization will fail if honest nodes did not cast their votes due to communication failure or dishonest votes cast by malicious actors \cite{xu2020raft}.
 
 Primary implementations of consensus in the private blockchain were based on Paxos over the last decade. Unfortunately, it is not easy to understand due to its intricate architecture, and aloof nature for practical systems. Therefore, a new consensus protocol (so-called RAFT) was developed that has a concise definition and comparable performance. RAFT reduces the degree of non-determinism by decomposing the nodes into two types of roles, one leader and the others are followers. RAFT implements the consensus by first electing a leader that receives log entries/transaction details from the client. The leader has the responsibility to manage the new log entries and replicate them on the servers. In contrast, followers are passive nodes that only respond to the request of the leader. The leader can only be replicate the logs on servers when the majority of followers successfully vote back to the leader \cite{ongaro2014search}.


Blockchain technology serves as securing different IoT applications, but it also vulnerable to different attacks \cite{ferrag2018blockchain}. These attacks of the blockchain systems lead to the failure of the consensus process responsible for verifying the transactions in the blockchain. \\
This work studies two types of active malicious attacks: jamming attacks and impersonation attacks in a RAFT based blockchain system. The coverage probability is derived in order to assess the impact of jamming in uplink and downlink transmissions. \\
On the other hand, physical layer authentication to counter impersonation attacks is an emerging domain in the field of information security \cite{aman2022security}.  Recently, to counter the impersonation attacks at the physical layer, \cite{Aman:Access:2018} and \cite{waqas:ICC:2020} propose distance, angle of arrival and position, \cite{Ammar:VTC:2017} and \cite{Aman:UCET:2019} propose channel impulse response, and  \cite{Aman:UCET:2020} exploits the lack of hardware reciprocity as features or device fingerprints to carried out authentication. We in this work encountered impersonation attacks by exploiting the pathloss of the transmitter nodes.

{\bf Contributions.} This work considers a RAFT-based IoT blockchain network that comprises a leader node and many follower nodes which together verify a blockchain transaction upon request from a blockchain client. To the best of authors' knowledge, this work is the first that considers two most prominent kinds of active attacks (jamming and impersonation) on the IoT blockchain networks. Specifically, two main contributions of this paper are as follows:
\begin{itemize}
    \item When the IoT blockchain network is under jamming attack, we utilize the stochastic geometry tool to derive the coverage probabilities for both uplink and downlink IoT transmissions, as a function of important system parameters, e.g., transmit power of legitimate and jamming nodes, intensity and relative geometry of the jamming nodes, etc. 
    \item When the IoT blockchain network is under impersonation attack, we propose a novel counter-method that enables a receive IoT node to exploit the pathloss of a transmit IoT node as its fingerprint to construct a two-step testing approach (i.e. Maximum Likelihood test followed by the binary hypothesis test). Furthermore, we provide closed-form expressions for the three error probabilities of interest, i.e., false alarm, missed detection and miss-classification.
\end{itemize}

{\bf Outline.} The rest of this paper is organized as follows. Section II provides a detailed description of the considered system model. Section III considers the scenario of a jamming attack on the IoT blockchain network, and derives the coverage probability for both uplink and downlink IoT transmissions. Section IV considers the scenario of an impersonation attack on the IoT blockchain network, and presents a novel countermeasure that enables a receive IoT node to exploit pathloss as fingerprint of a transmit IoT node to implement a hypothesis testing test for transmitter identification. Section V provides simulation results. Finally, Section VI concludes the paper.

{\bf Notations.} Unless specified otherwise, $\vert.\vert$ and $\Vert.\Vert$  denote the modulus and the 2-norm respectively, $\mathbb{E}(.)$ is the expectation operator, boldface letters such as $\mathbf{X}$ represents a vector and $\mathcal{CN}$ means complex normal.
\begin{figure}[htp]
    \centering
    \includegraphics[width=8cm]{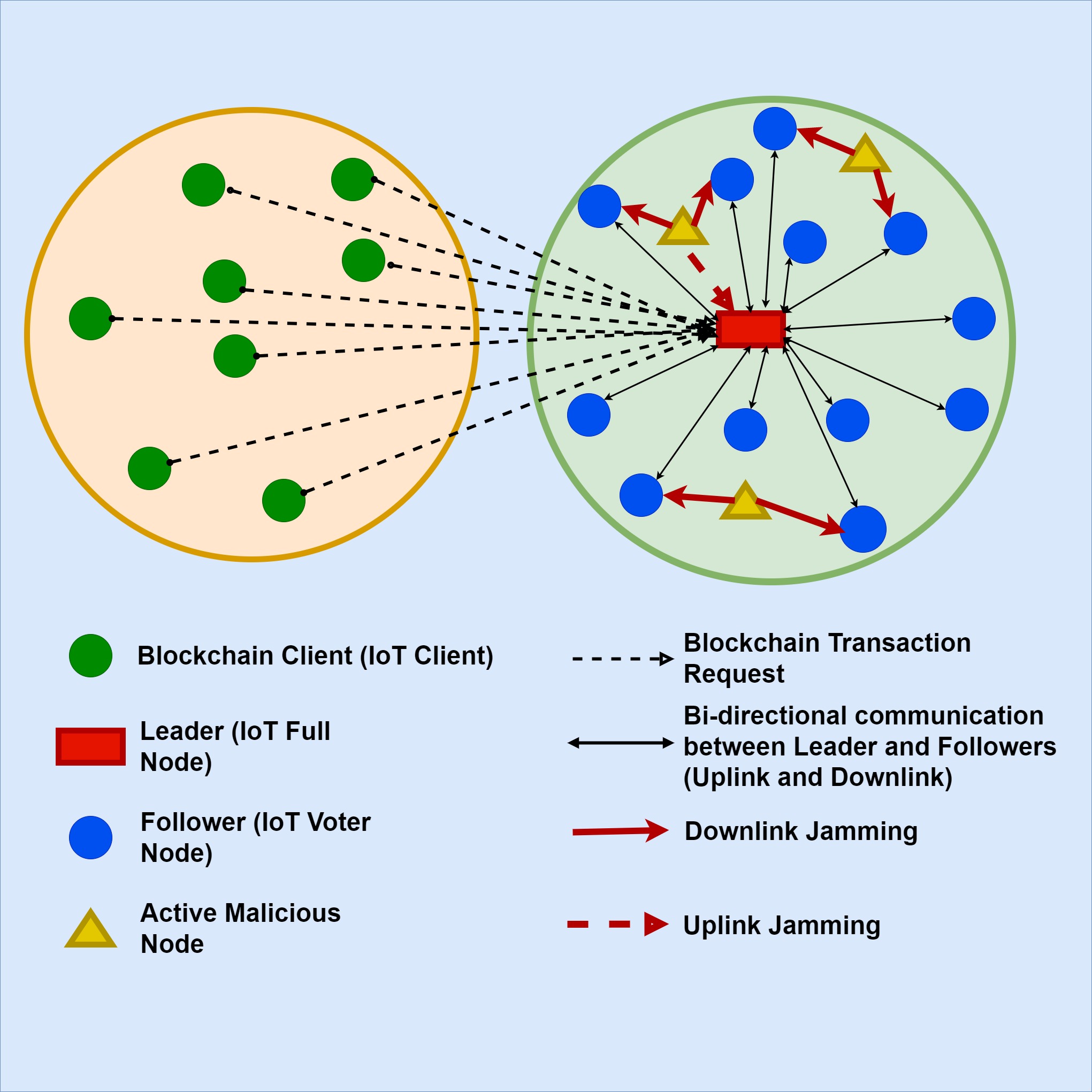}
    \caption{System Model}
    \label{fig:sys}
\end{figure}

%% file: sec2.tex
 \section{System Model \& Background}
\label{sec:sys-model}

\subsection{System Model}	
The IoT blockchain network based on the RAFT consensus algorithm comprises two parts, wireless consensus network and IoT clients, as shown in Fig. \ref{fig:sys}. The two parts may or may not be geographically isolated. The figure illustrates the communication
network topology and the IoT nodes' roles interchangeably in different business models. Any IoT node in the network can play a client's role that sends out transaction-requests or a leader/follower in the consensus process. The followers and malicious nodes locations in a 2-Dimensional free space are modeled as Poisson Point Process (PPP)  and leader (full node) is fixed at geo-center of consensus network. We consider that malicious nodes are  active malicious nodes who are capable of launching jamming and impersonation attacks. We assume that malicious nodes either launch jamming attacks or impersonation attacks in a given time. 

\subsection{RAFT Consensus Protocol}
RAFT consensus algorithm begins by first electing a leader who receives log entries/transaction-details from the client. The leader has high capabilities such as better performance and reliability to manage the new log entries and replicate them into the blockchain. In contrast, followers are passive nodes that only respond to the request of the leader. The leader can only insert the transaction-details into the blockchain when most followers successfully vote back to the leader.
The RAFT consensus mechanism is triggered as follows. Firstly, the leader receives the transaction information from clients and performs some necessary actions to form a block. Secondly, a leader will communicate with follower nodes via a downlink (DL) broadcasting channel to approve the block. When followers successfully received a DL message, they verifying and send the voting message on the multi-access uplink (UL) channel for confirmation. Lastly, a leader will count the vote to achieve consensus; if the leader gains the majority more than 50\%  from the followers (voters).

This paper focuses on communication inside the consensus network, on both DL and UL.

 \section{Jamming Attack on IoT Blockchain Network}
\label{sec:Jamming}

In the RAFT consensus algorithm, a client transmits transaction-requests to the leader node to make a consent with the followers by considering that all follower nodes are honest. Therefore, the threat of failure among the nodes is due to communication faults. Communication failure is due to the opposing jammers that can exist in the network due to the openness of the wireless channel.
The leader would lose the vote(follower) if communication links (DL or UL) failed due to opposing jammers and channel fading. This section finds success/coverage probability defined as receiving Signal to Interference and Noise Ratio (SINR) is greater than its predefined threshold between a leader and associated follower. All followers operate in the same frequency band, so they cause increased interference that ultimately degrades the received SINR and lowers the success probability. Different protocols were employed in the private blockchain networks to avoid interference and collision among the follower nodes, such as centralized radio resource allocation and Carrier Sense Multiple Access (CSMA) or transmission interval made large enough so that collision is negligible.

\subsection{Coverage Probability for IoT Transmissions on the Downlink}
\label{sec:Downlink}

When the leader transmits, then a typical follower receives the following baseband signal:
\begin{equation} 
\label{eq:y}
y = \sqrt{P} Hs + \sum_{j\in\phi_{J}} \sqrt{P_j} H_j s + n,
\end{equation}

where $P$ ($P_j$) is the transmit power of the leader ($j$-th jammer) node, $s$ is the transmitted symbol, $H=h/\sqrt{R^\alpha}$ ($H_j=h_j/\sqrt{R_j^\alpha}$) represents the wireless propagation from leader (jammer) to the follower, and $R$ ($R_j$) is the random distance between the leader (jammer) and the follower node. Further, for the leader-follower link, $R^{-\alpha}$ is the large-scale fading/pathloss component, ${\alpha}$ is the pathloss exponent, $h \sim \mathcal{CN}(0,1)$ is the small-scale fading component. Moreover, $\sqrt{I_{J}} =\sum_{j\in\phi_{J}} \sqrt{P_j} H_j s$ is the aggregate interference amplitude due to multiple jammers where $\phi_{J}$ indicates that the jammers are distributed in a 2-dimensional free-space as PPP. Finally, $n \sim \mathcal{CN}(0,\sigma^2)$ is the additive white Gaussian noise with power $\sigma^2$.

We consider an interference-limited scenario (i.e., the interference is much larger than the noise). This allows us to consider the signal-to-interference ratio (SIR) as the performance metric. The SIR of a typical follower associated with the leader node is given as:  

\begin{equation}
\label{eq:SIR}
SIR^{DL}=\dfrac{P|h|^2{R^{-\alpha}}}{I_J},
\end{equation}
where $I_{J}=\sum_{j\in\phi_{J}}{P_{j}}|h_j|^2{\Vert\mathbf{X}_j\Vert^{-\alpha}}$ is the aggregate interference power, $h_j\sim \mathcal{CN}(0,1)$ is the small-scale fading component on the jammer-follower channel, and $\mathbf{X}_j$ is the random location of $j$-th jammer or Poisson point (note that $\Vert\mathbf{X}_j\Vert=R_j$). 
	
The IoT transmission from the leader to any given follower node on the DL will be considered successful only when the received SIR is greater than a pre-specified threshold $\beta_D$ \cite{lu2021stochastic}. Thus, the transmission success probability, or, the coverage probability for the DL ($\mathcal{P}^{DL}_c$) is defined as follows:
\begin{equation}
\label{eq:hh}
    \mathcal{P}^{DL}_c(\alpha,\beta_D)= \mathcal{P}\biggl[SIR^{DL}>\beta_D\biggl]=\mathcal{P}\biggl[\dfrac{P|h|^2{R^{-\alpha}}}{I_{J}}>\beta_D\biggl].
\end{equation}

Now, we assume that a typical follower is at distance $r$ from the leader, then the coverage probability can be expressed as:
\begin{equation}
\label{eq:Pdl}
    \mathcal{P}^{DL}_c(\alpha,\beta_D)= \mathbb{E}_{R} \biggl[\mathcal{P}\biggl[SIR^{DL}>\beta_D \mid R=r\biggl]\biggl],
\end{equation}
\begin{equation}
=\int_{r>0}^\infty \mathcal{P}\biggl[SIR^{DL}>\beta_D \mid r\biggl]f_R(r)dr, \nonumber
\end{equation}
where $f_R(r)$ is the probability density function (PDF) of $R$, and  is given as \cite{Mathai1999AnIT} \footnote{We consider the elected leader node at the origin of considered space}: 
\begin{align}
f_R(r)=2\pi\rho_T r \exp({-\rho_T \pi r^2}), 
\end{align}
where $\rho_T$ is the intensity/density of the IoT nodes. At this stage, first we need to compute $\mathcal{P}\big[SIR^{DL}>\beta_D \mid r\big]$ which can be expressed as:
\begin{equation}
\mathcal{P}\biggl[SIR^{DL}>\beta_D \mid r\biggl] = \mathcal{P}\biggl[|h|^2>\dfrac{{r^{\alpha}}\beta_D }{P} I_{J}\biggl]. 
\end{equation}
As $|h|^2\sim \exp(1)$, we can write: 
\begin{equation}
\mathcal{P}\biggl[|h|^2>\dfrac{{r^{\alpha}}\beta_D }{P} I_{J}\biggl] = \mathbb{E}_{I_J}\biggl[\exp(-\dfrac{{r^{\alpha}}\beta_D }{P} I_{J})\biggl]=\mathcal{L}_{I_{J}}(\dfrac{{r^{\alpha}}\beta_D }{P}),
\end{equation}
    where $\mathcal{L}_{{I}_{J}}(s)$ denotes the Laplace transform of the aggregate interference $I_{J}$ which is computed in Appendix A (with variable $s=\dfrac{{r^{\alpha}}\beta_D }{P}$).



Putting back the result of Appendix A to Eq. \ref{eq:Pdl} we get the following final expression of coverage probability for DL: 


\begin{equation}
\begin{split}
& \mathcal{P}^{DL}_c(\alpha,\beta_D)={2\pi\rho_{T}}\bigintsss_{r\geq 0}\exp\biggl(\dfrac{\pi \rho_J \gamma_{j}\beta_D r^{\alpha}}{(\alpha/2)-1} 
    \biggl[z_2^{(2-\alpha)}\\
    &{}_{2}F_{1}(1,1-{\dfrac{2}{\alpha}},2-{\dfrac{2}{\alpha}},-\gamma_J \beta_D (\dfrac{r}{z_2})^\alpha)-z_1^{(2-\alpha)}\\
    &{}_{2}F_{1}(1,1-{\dfrac{2}{\alpha}},2-{\dfrac{2}{\alpha}},-\gamma_J \beta_D         (\dfrac{r}{z_1})^\alpha)\biggl]-\rho_T \pi r^2 \biggl) r dr.
\end{split}
\end{equation}


 \subsection{Coverage Probability for IoT Transmissions on the Uplink}
\label{sec:Uplink}
To achieve the consensus, followers send the voting message on the multi-access UL channel for confirmation after receiving a DL message. Consensus will be achieved if more than  $50\%$  from the followers successfully verify the transaction on the UL channel. So, we need to compute the success probability on the UL channel. We assume that CSMA is the medium access technique adopted by the followers and transmissions on UL is available all the time (i.e., no idle channel). Then the coverage probability for a typical follower node on the UL is given as  \cite{lu2021stochastic}:
\begin{equation}
\mathcal{P}^{UL}_c(\alpha,\beta_U)=\mathcal{P}\biggl[SIR^{UL}>\beta_U\biggl]=\mathcal{P}\biggl[|h^{U}|^2>\dfrac{{R_{U}^{\alpha}}\beta_U }{P_{F}} I_{J}^{U}\biggl],
\end{equation}
where $\beta_U$ is a predefined threshold for UL, $h^{U}$ is the channel gain and $R_{U}$ is the distance from a typical follower to the leader on UL, $P_F$ is the transmit power of a typical follower,  and  $I_{J}^{U}$ is the aggregated interference to the leader.
CSMA makes sure no interference from the other follower nodes on the UL and hence the interference/jamming is due to the jammers only. This makes the coverage probability formulation the same as we do for DL.
We compute UL success probability using the same procedure as above, we get:

\begin{equation}
\begin{split}
&   \mathcal{P}^{UL}_c(\alpha,\beta_U)={2\pi\rho_{T}}\bigintsss_{r\geq 0}\exp\biggl(\dfrac{\pi \rho_J \gamma_{j}^U\beta_U r_{U}^{\alpha}}{(\alpha/2)-1} 
    \biggl[z_2^{(2-\alpha)}\\
    &{}_{2}F_{1}(1,1-{\dfrac{2}{\alpha}},2-{\dfrac{2}{\alpha}},-\gamma_j^U \beta_U (\dfrac{r_{U}}{z_2})^\alpha)-z_1^{(2-\alpha)}\\
    &{}_{2}F_{1}(1,1-{\dfrac{2}{\alpha}},2-{\dfrac{2}{\alpha}},-\gamma_j^U \beta_U         (\dfrac{r_{U}}{z_1})^\alpha)\biggl]-\rho_T \pi r_{U}^2 \biggl) r_{U} dr_{U},
\end{split}
\end{equation}
where $\rho_J$ is the intensity of jammer nodes and $\gamma_{j}^U ={P_{j}}/{P_{F}}$.

\subsection{Overall Coverage Probability}
\label{sec:joint}
Consensus is achieved when followers successfully receive the leader's request to verify the transaction-detail over the DL channel and respond to the leader over the UL channel. So, we derive the joint coverage probability in an IoT blockchain network as follows
\cite{7343565}:

\begin{equation}
    \mathcal{P}_{c}={\mathcal{P}[SIR^{UL}>\beta_U]}.{\mathcal{P}[SIR^{DL}>\beta_D]}
\end{equation}

\section{Impersonation Attack on IoT Blockchain Network}

In the RAFT consensus algorithm, due to broadcast nature of wireless communication, the consensus algorithm may be failed due to the impersonation attacks launched by nearby malicious/illegitimate node(s) (so called Eve(s)). In the impersonation attacks, malicious nodes try to claim themselves as legitimate nodes/followers by utilizing a forged character in order to destroy the consensus mechanism. \\
In this work, we provide a physical layer authentication mechanism to counter the impersonation by illegitimate nodes. 
We, in this work exploit the pathloss of the transmitter node as device fingerprint to counter the impersonation.  
 We assume a realization\footnote{Typically, in physical layer authentication, one needs to know the exact number of transmitting nodes in order to evaluate its performance. Therefore, in this part, we take a single realization of PPP in order to fix the total number of nodes (both, malicious and legitimate).} of PPP , specifically, $M$ follower (legitimate) nodes $\{F_i\}^{M}_{i=1}$ and $N$ Eve (malicious/illegitimate) nodes $\{E_j\}^N_{j=1}$ are considered in  a 2-Dimensional space, and a leader is placed at the center of the considered region. We assume that the transmitter nodes transmit with a fixed transmit power so that the leader can compute the pathloss. We also assume that the malicious nodes are transmitting with the same transmit power in order to stay stealth in the region \footnote{High power transmission will easily identify the transmitter as malicious node. To get success in impersonating the legitimated nodes, malicious nodes transmit with same power as legitimate nodes do}. The pathloss $\Psi$ in dB of a transmitter at the distance $d$ from receiver is given as:

\begin{align}
\label{eq:THzPL}
\Psi[dB]=10 \alpha \log_{10}(d),
\end{align}

where $\alpha$ is the pathloss exponent.

\subsection{The Proposed Authentication Method}
We assume that CSMA is the approach used by the followers and malicious nodes to cast their votes. We assume that malicious node $E_j$ could cast a vote, pretending to be a legitimate follower node when the channel is completely idle by the followers and hence, no collision. The leader is supposed to authenticate each received casting vote and correctly achieved the consensus. Furthermore, we assume that the leader already have the ground truths of legitimate nodes which he gets via prior training on a secure channel. The ground truth vector can be denoted by $\mathbf{\Psi} = \{\Psi_1, ..., \Psi_M\}^T$. As discussed earlier, we will authenticate the transmitter based on the pathloss feature. So the noisy measurement of pathloss $z = \Psi + n$ at a given time-slot is obtained by using the pulse-based method, where $\Psi$ is the pathloss and $n \sim \mathcal{N}(0,\sigma^2)\,$ is the noise/estimation error.  To counter the impersonation by malicious nodes, we first do Maximum Likelihood (ML) test as follows: 

\begin{equation} 
\label{eq:ML}
i^* = \underset{i}{\max} \quad f(z\mid \Psi_i),
\end{equation}
where $f(z\mid \Psi_i)$ is the likelihood function or conditional PDF.\\
Equivalently, we can write \ref{eq:ML} as:

\begin{equation} 
\label{eq:ML-pl}
(TS^*,i^*) = \underset{i}{\min} \quad |z-\Psi_i|,
\end{equation}

where $TS^*$ is the minimum value of test statistics amd $i^*$ returns the index of transmitter node which is decided through ML. Next, we decide for impersonation through binary hypothesis testing as follow:

\begin{equation}
	\label{eq:H0H1}
	 \begin{cases} H_0 (\text{no impersonation}): & TS^*=\underset{i}{\min} |z(t)-\Psi_i| < \epsilon \\ 
                   H_1 (\text{impersonation}): & TS^*=\underset{i}{\min} |z(t)-\Psi_i| > \epsilon \end{cases},
\end{equation}
where $\epsilon$ is a small test threshold and is a design parameter that decides a vote from the node is accepted or not. 
Equivalently, we have:
\begin{align} 
\label{eq:bht}
{TS}^* \gtrless_{H_0}^{H_1} {\epsilon}.
\end{align}

The hypothesis
$H_0$ inferred that the legitimate node transmits the vote. Alternatively, the hypothesis $H_1$ implies that a illegitimate node transmits a vote. 
Further, we present closed form expressions for the error probabilities. We have three types of errors resulted from the above tests. These errors are: false alarm, missed detection and miss-classification. The probability of false alarm is the probability that a legitimate follower casts a vote, but the leader notices it as illegitimate, denoted as $\mathcal{P}_{fa}$. The Probability of missed detection  is the probability that a impersonate node cast a vote, but the leader considered it as a legitimate vote, denoted as $\mathcal{P}_{md}$. Last, the probability of miss-calssification is, when no impersonation is detected but a wrong transmitter node is decided.

We, in this work follow the Neyman-Pearson lemma \cite{923720} where $\epsilon$ test threshold for a pre-defined false alarm $\mathcal{P}_{fa}$ can be chosen such that missed detection probability $\mathcal{P}_{md}$ is minimized. Thus, the error probabilities for the above hypothesis tests are:
\begin{equation}
\begin{split}
\mathcal{P}_{fa} &= \mathcal{P}(H_1|H_0)= \sum_{i=1}^M \mathcal{P}(TS^*>\epsilon|F_i)\pi (i) \\
&=\sum_{i=1}^M 2Q(\frac{\epsilon}{\sigma})\pi (i)=2Q(\frac{\epsilon}{\sigma})\sum_{i=1}^M \pi (i)=2Q(\frac{\epsilon}{\sigma})
\end{split}
\end{equation}
Where $\pi (i)$ is the prior probability of of legal node  $F_i$ (we consider equal priors in our work), and $Q(x)=\frac{1}{\sqrt2\pi}\int_x^\infty e^{\frac{-t^2}{2}}dt$ is the complementary CDF of a standard normal distribution  Thus, the threshold could be computed as follows:

\begin{equation}
\label{eq:epsi}
\epsilon = \sigma Q^{-1}(\frac{\mathcal{P}_{fa}}{2})
\end{equation}

\subsection{The Performance of the Proposed Method}

Next, $\mathcal{P}_{md}$ is given as:
\begin{equation}
\begin{split}
&{\mathcal{P}}_{md}  = \mathcal{P}(H_0|H_1) = \mathcal{P}(TS^*<\epsilon|E_j) \\
& =\sum_{j=1}^{N} \sum_{i=1}^{M} \bigg[ Q(\frac{\Psi_i - \Psi_j - \epsilon}{\sigma}) - Q(\frac{\Psi_i - \Psi_j + \epsilon}{\sigma}) \bigg] \frac{\pi (j)}{M}
\end{split}
\end{equation}
where $\pi (j)$ is the prior probability of impersonator node $E_j$. 

Since the probability of missed detection is a random variable, so the expected value $\bar{\mathcal{P}}_{md}:=\mathbb{E}(\mathcal{P}_{MD})$ is as follows:
\begin{equation} 
\begin{split}
&\bar{\mathcal{P}}_{md}  =\sum_{j=1}^N \frac{1}{\Delta}\pi (j). \\
& \resizebox{.5 \textwidth}{!}{$\bigg( \bigintsss_{\Psi_{min}}^{\Psi_{max}} \sum_{i=1}^M Q(\frac{\Psi_i - \Psi_j^{(E)} - \epsilon}{\sigma})  - Q(\frac{\Psi_i - \Psi_j^{(E)} + \epsilon}{\sigma}) d\Psi_j^{(E)} \bigg)$} \\
& =\sum_{j=1}^N \frac{1}{\Delta}\pi (j). \\
&\resizebox{0.5 \textwidth}{!}{$\bigg( \bigintsss_{\Psi_{min}}^{\Psi_{max}} \sum_{i=1}^M Q(\frac{\Psi_i - \Psi^{(E)} - \epsilon}{\sigma}) - Q(\frac{\Psi_i - \Psi^{(E)} + \epsilon}{\sigma}) d\Psi^{(E)} \bigg)$}
\end{split}
\end{equation}
where we have assumed that the unknown pathloss $\Psi_j\sim U(\Psi_{min},\Psi_{max})$ $\forall j$, and $\Delta=\Psi_{max}-\Psi_{min}$. 

Now, we investigate the authentication of the casted vote  by identifying transmitter identity using the ML based approach. 
The error probability of miss-classified node $\mathcal{P}_{mc}$ resulting from Eq. \ref{eq:ML-pl} is given as:    
\begin{equation}
\mathcal{P}_{mc} = \sum_{i=1}^M \mathcal{P}_{mc|i}.\pi(i)
\end{equation}
where $\mathcal{P}_{mc|i}$ is the probability that leader notice that the vote is casted by follower $F_j$ but the vote is actualy casted by follower $F_i$ where $i\neq j$.  For the hypothesis test of (\ref{eq:bht}), $\mathcal{P}_{mc|i}$ is given as:
\begin{equation}
\label{eq:pmc}
\mathcal{P}_{mc|i}=1-\bigg( Q(\frac{\tilde{\Psi}_{l,i}-\tilde{\Psi}_i}{\sigma}) - Q(\frac{\tilde{\Psi}_{u,i}-\tilde{\Psi}_i}{\sigma}) \bigg)
\end{equation}
where $\tilde{\Psi}_{l,i}=\frac{\tilde{\Psi}_{i-1}+\tilde{\Psi}_i}{2}$, $\tilde{\Psi}_{u,i}=\frac{\tilde{\Psi}_{i}+\tilde{\Psi}_{i+1}}{2}$. Additionally, $\mathbf{\tilde{\Psi}}=\{\tilde{\Psi}_{1},...,\tilde{\Psi}_{M}\}=\text{sort}(\mathbf{\Psi})$ where sort operation (.) sorts a vector in an increasing order. For the boundary cases, e.g., $i=1, i=M$, $\tilde{\Psi}_{l,1}=\Psi_{min}$, $\tilde{\Psi}_{l,M}=\Psi_{max}$ respectively.
\\ Note that the proposed mechanism will be executed at the leader node, which becomes uplink transmission according to the discussion of Section \ref{sec:Jamming}. For the downlink, the mechanism will be executed at followers having a single ground truth of the leader node. In this case, we will have two error probabilities (i.e., false alarm and missed detection) with a kind of similar error expressions given above (i.e., except summation for $N$ legitimate nodes). We omit the discussion of it for the sake of brevity.

%% file: sec3.tex
 \section{Results And Discussions}
 \label{sec:RESULTS}
We use MATLAB $2019a$ for simulations. The important simulation parameters for generated figures are mentioned in TABLE \ref{tb:1}, unless otherwise stated. 

 \begin{table}[ht]
 \begin{center}
      \begin{tabular}{ |p{4cm}|p{4cm}| }
 \hline
 Parameters& Configuration \\
 \hline
 $P_L$, $P_F$, $P_j$   & 30 dBm, 20 dBm, 10 dBm     \\
  \hline
 Pathloss exponent $\alpha$ &   $3$ \\
 \hline
  $\rho_T$, &  $15/\pi(500)^2 $  \\
 \hline
  \hline
Area  & $\pi  (500m)^2$     \\
  \hline
  $M$, $N$  & $5$, $5$     \\
  \hline
\end{tabular}
 \end{center}
 \caption{Important simulation parameters}
 \label{tb:1}
  \end{table}
	\subsection{Coverage probability performance of the IoT blockchain network under jamming attack}
	
	Under jamming attack, both legitimate and malicious nodes are deployed according to the PPP with $\rho_T$ intensity for legitimate nodes, and $\rho_J$ intensity for jammer/malicious nodes. The leader is placed at the origin. 
	
\begin{figure}[htp]
    \centering
    \includegraphics[width=9cm]{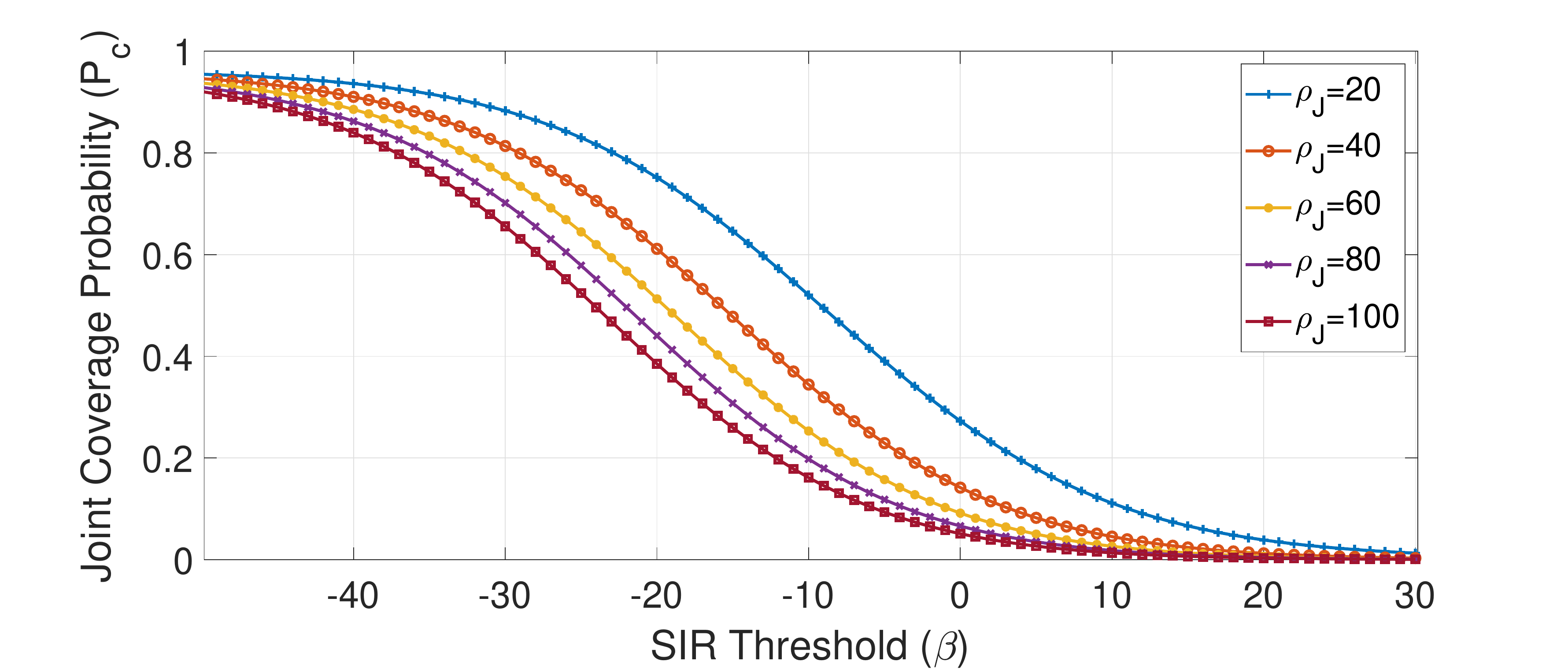}
    \caption{Transaction success probability vs SIR threshold $\beta$ for different jammers densities }
    \label{fig:F1}
\end{figure}

	Fig. \ref{fig:F1} presents the joint coverage probability for the  different SIR threshold values of $\beta$ (in dB) and intensity of jammers $\rho_J$. We observe that the transaction success rate (i.e., $\mathcal{P}_c$) declines as the SIR threshold increases. Fig. \ref{fig:F1} also reveals that increasing the jammer intensity severely degrades the transaction success rate.

\begin{figure}[htp]
    \centering
    \includegraphics[width=9cm]{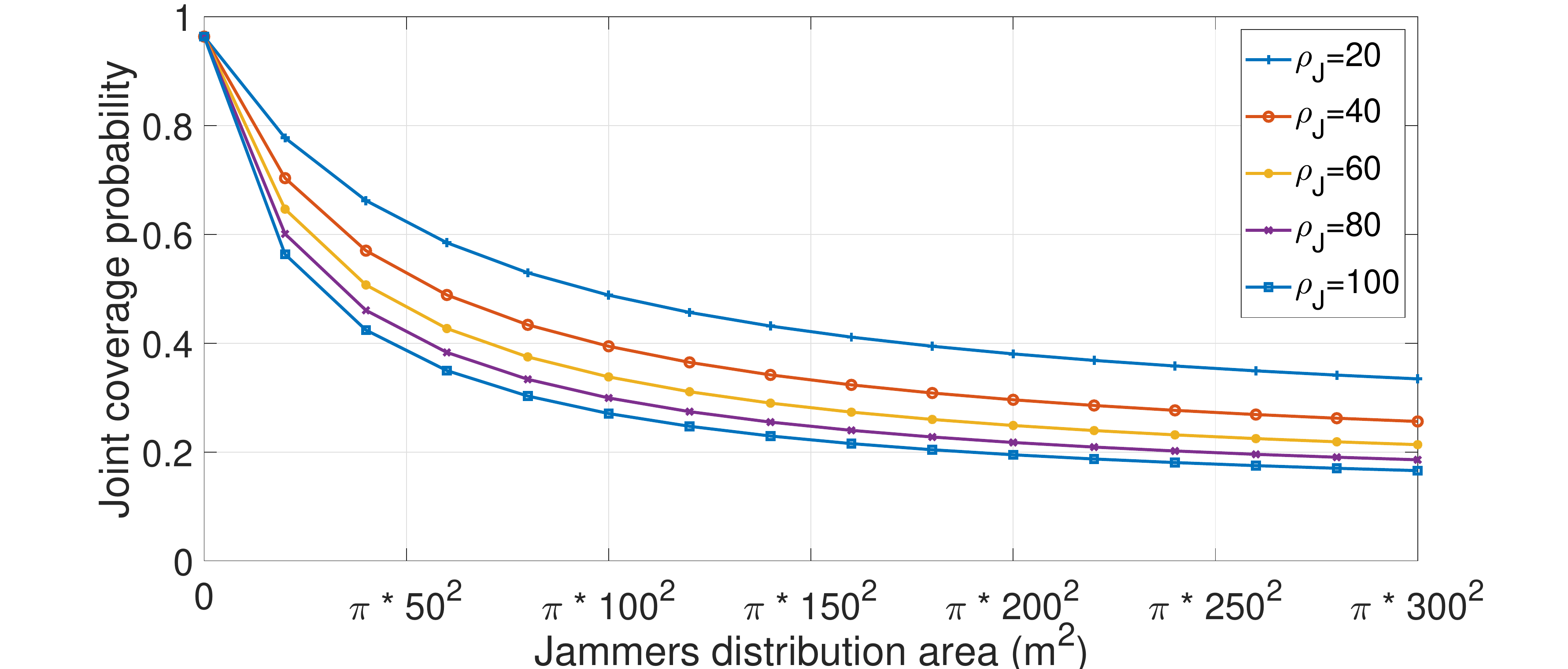}
    \caption{Transaction success probability vs effective jamming area  }
    \label{fig:F2}
\end{figure}

Fig. \ref{fig:F2} demonstrates the behaviour of joint coverage probability against different radius values (i.e., effective area) of distributed jammers. Specifically, we set $\rho_T=\rho_J$, $z_1=0$ and vary $z_2$ from $0$ to $300$ by step of $20$. We observe that the increase in the effective jamming area of jammers causes low SIR values that lowers the transaction success rate. We can also see the degradation in the performance of success rate with increase in intensity of jammers $\rho_J$.

\begin{figure}[htp]
    \centering
    \includegraphics[width=9cm]{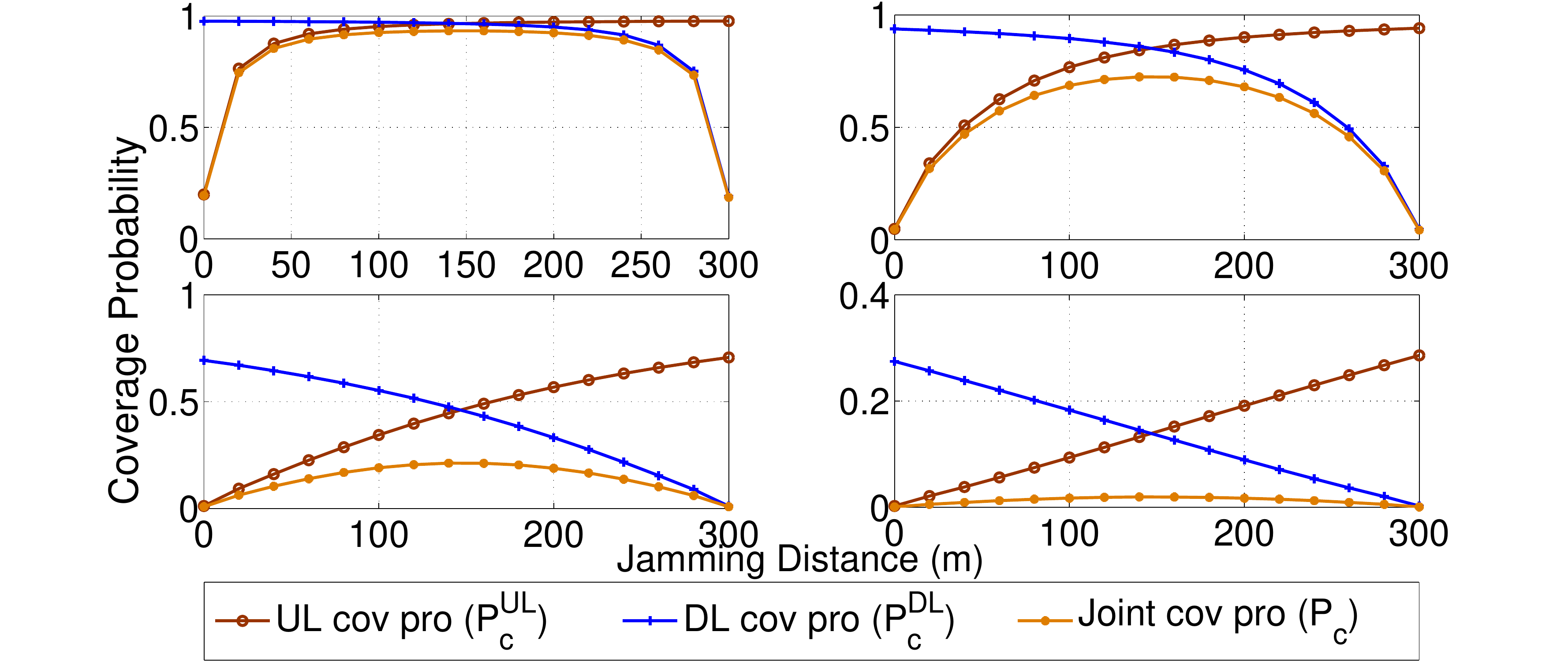}
    \caption{Transaction success probability vs jamming distance}
    \label{fig:F3}
\end{figure}

Fig. \ref{fig:F3} presents the three computed coverage probabilities: UL, DL, and joints coverage probabilities against  the jamming distance (i.e., $z_1$) from the origin or leader. We vary $z_1$ from $0m$ to $300m$ and keep $z_2=z_1+50m$. In the upper two plots, the $\beta$ vary from $-30$ dB to $-20$ dB from left to right, and from $-10$ dB to $0$ dB in the lower two plots.  We notice that the effective jamming area (or can be thought as jammers) moves away from the leader, making the interference lower at the leader and resulting in higher UL coverage probability. Followers are deployed in $\pi (500m)^2$ area, so moving of jamming distance from the origin produces low DL coverage probability as jammers are getting closed to followers. On the other hand, joint coverage probability first enhances and then goes down. Note that we need coverage probabilities greater than the $0.5$ to achieve the consensus.

\subsection{Authentication performance of the IoT blockchain network under impersonation attack}
\begin{figure}[htp]
    \centering
    \includegraphics[width=9cm,height=8cm]{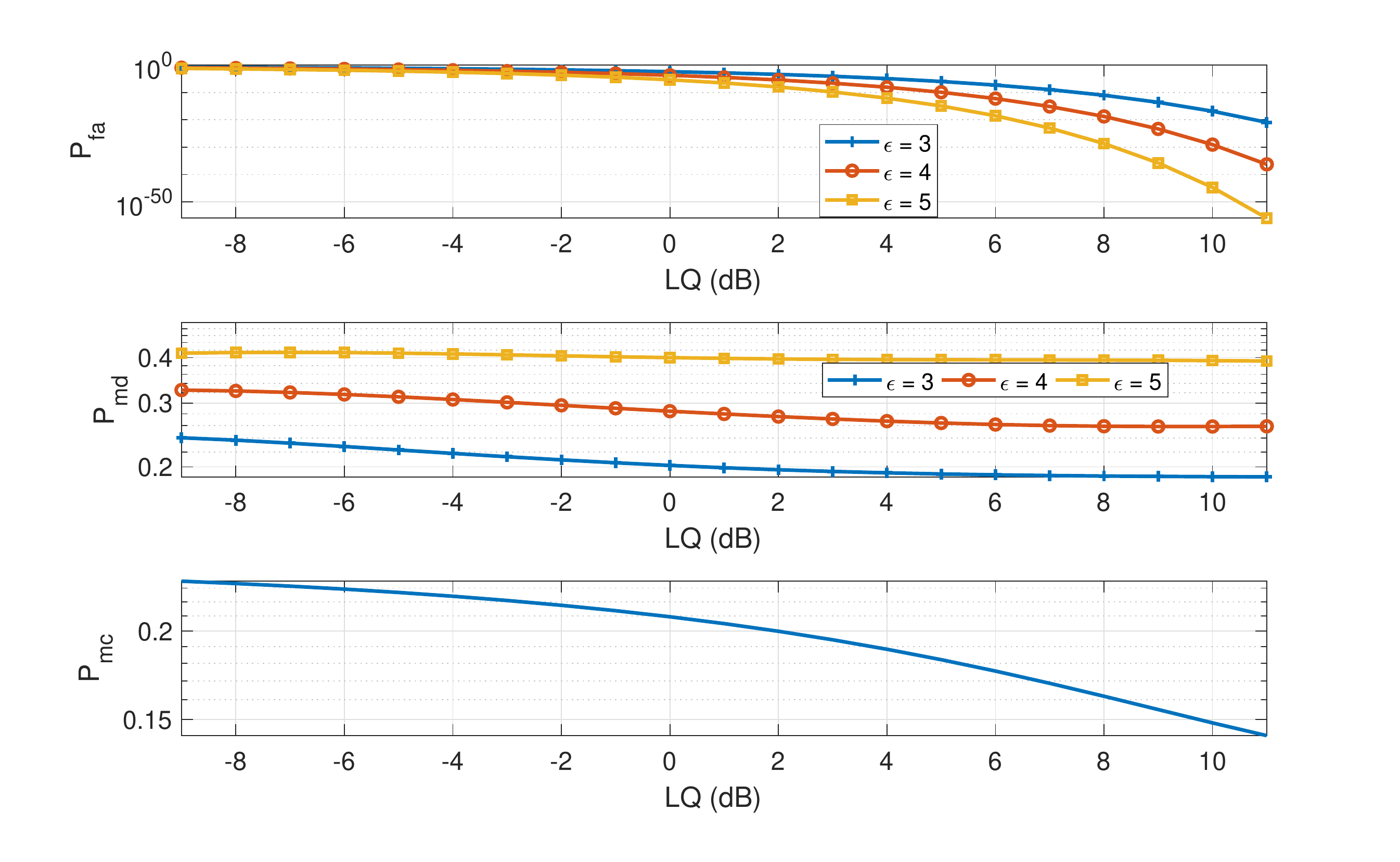}
    \caption{Error Probabilities against $LQ$. From top to bottom: Probability of false alarm $P_{fa}$, probability of missed detection $P_{md}$ and probability of misclassification $P_{mc}$. }
    \label{fig:F5}
\end{figure}

We choose a realization of PPP for malicious and legitimate nodes where we have $M=N=5$ number of malicious and legitimate nodes. We set the link quality  as $LQ=1/\sigma^2$, which means that more the uncertainty in the estimation/noise implies poor link quality and vice versa. 
We plot the error probabilities (false alarm, missed detection and miss classification) as  functions of  $LQ$ (in dB) in Fig. \ref{fig:F5}.  We observe from this figure that pathloss can be exploited to counter the impersonation attacks in wireless blockchains networks. A designed parameter $\epsilon$ can be set for the desired level of security. We observe that we can not minimize both errors for a given link quality simultaneously. In other words, increasing $\epsilon$ improves false alarm but degrades the missed detection. The lower plot  of Fig. \ref{fig:F5} shows the miss classification error against link quality and it demonstrates that it is not a  function of   $\epsilon$, and that's why a single curve for all the three given choices of $\epsilon$.

\begin{figure}[htp]
    \centering
    \includegraphics[width=9cm,height=5cm]{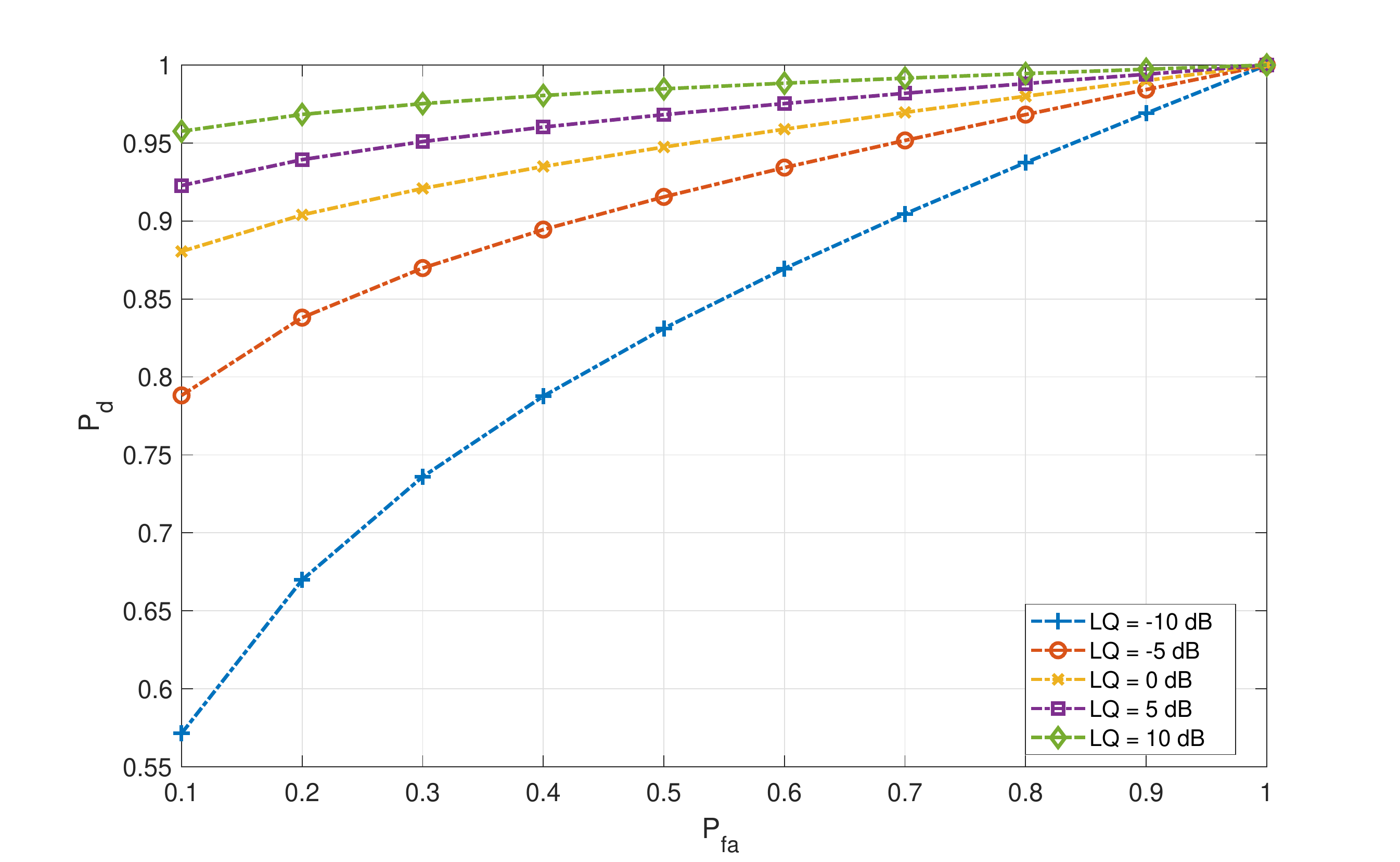}
    \caption{Receiver Operating Characteristic (ROC) curves: probability of detection $P_d$ can be set to a desired level while compromising on probability of false alarm $P_{fa}$  }
    \label{fig:F6}
\end{figure}

Figure \ref{fig:F6}
shows the Receiver Operating Characteristic (ROC) curves comprises two
error probabilities ($P_d$ and $P_{fa}$). where $P_d = 1 - P_{md}$ is the detection
probability defined as the probability of correctly deciding malicious nodes . We sweep $P_{fa}$ from zero to one, and find $P_{md}$ for different values of $1/\sigma^2$. As we can see improvements in link quality improves $P_d$. These curves can be used to set the system to a desired level of security where for a given link quality, one can find a value of false alarm for a desired value of detection.

%% file: conclusion.tex
\section{Conclusion}
\label{sec:conclusion}
This paper studied active (jamming and impersonation) attacks on a RAFT-based IoT blockchain network. The impact of the jamming attack on the IoT blockchain network was evaluated via coverage probability analysis for both uplink and downlink IoT transmissions. On the other hand, the impersonation attack on the IoT blockchain network was countered by means of a novel, physical-layer method that exploited the pathloss of a transmit IoT node as its fingerprint to construct a binary hypothesis test for transmit node identification. To this end, closed-form expressions were provided for the probabilities of false alarm, missed detection and miss classification. Simulation results showed that for the jamming attack, an increase in the threshold value reduces the coverage probability and high intensity of jammers produces low coverage probability, while for the impersonation attack, pathloss can be used as device fingerprint and above $95 \%$ of detection probability can be achieved  with minimum of $0.1$ false alarm for a $10$ dB link quality.

%% file: appendix.tex
\section{Computation of $\mathcal{L}_{I_{J}}(s)$ for IoT transmission on the downlink}

The Laplace transform of the cumulative interference ${I_J}$ is defined as:

\begin{align}
\mathcal{L}_{I_{J}}(s)=\mathbb{E}_{I_{J}}\biggl[\exp(-s I_{J})\biggl],
\end{align}

\begin{align}
\mathcal{L}_{I_{J}}(s)=\mathbb{E}_{\phi_j,\{|h_j|^2\}}\biggl[\exp\biggl(-s {\sum_{j\in\phi_{J}}{P_{j}}|h_j|^2{\Vert\mathbf{X}_j\Vert^{-\alpha}}}\biggl)\biggl].   
\end{align}

Next, we use the property of exponential function that sum of exponential powers is the product of exponentials and  put the value of $s$ back in above expression to get: 

\begin{align}
\mathcal{L}_{I_{J}}(s)=\mathbb{E}_{\phi_j,\{|h_j|^2\}}\biggl[\prod_{{j\in\phi_{J}}}exp\biggl(- |h_j|^2\biggl(\dfrac{P_{j}}{P}\biggl) {\beta_D } {\Vert\mathbf{X}_j\Vert^{-\alpha}}{r^{\alpha}}\biggl)\biggl].
\end{align}
Let $\gamma_{j} ={P_{j}}/{P}$, as $\phi_j$ is independent with $|h_j|^2$ we can take one expectation (i.e., $\mathbb{E}_{\{|h_j|^2\}}$) inside, which is given below:

\begin{align}
\label{eq:PGFL}
\mathcal{L}_{I_{J}}(s)=\mathbb{E}_{\phi_j}[\prod_{{j\in\phi_{J}}}\mathbb{E}_{\{|h_j|^2\}}[exp(- |h_j|^2\gamma_{j} {\beta_D } (\dfrac{\Vert\mathbf{X}_j\Vert}{r})^{-\alpha})]].
\end{align}

Indeed, Eq. \ref{eq:PGFL} is Probability Generating FunctionaL (PGFL) of PPP, which can be expressed as:

\begin{align}
&\mathcal{L}_{I_{J}}(s)= \\& \exp\biggl(-\rho_J \bigintsss_{\mathcal{A}}1-\mathbb{E}_{|h|^2}[exp(- |h|^2\gamma_{j} {\beta_D } (\dfrac{\Vert\mathbf{x}\Vert}{r})^{-\alpha})] d\mathbf{x}\biggl), \nonumber
\end{align}
where $\rho_J$ is the intensity of jammer nodes, $\mathcal{A}$ is the effective 2D area where jammers signals are prominent or can affect the transmissions.
Now, converting $x$ into polar form as $x=(r_j, \theta)$ ($j$ subscript is used in order to differentiate it from the earlier used $r$ (distance of follower)), we have,
\begin{align}
&\mathcal{L}_{I_{J}}(s)= \\& \exp\biggl(-2\pi\rho_J \bigintsss_{z_1}^{z_2}\biggl(1-\mathbb{E}_{|h|^2}[exp(- |h|^2\gamma_{j} {\beta_D } (\dfrac{{r_j}}{r})^{-\alpha})]\biggl) r_j dr_j\biggl), \nonumber
\end{align}
 which can written as:

\begin{align}
\mathcal{L}_{I_{J}}(s)=\exp\biggl(-2\pi\rho_J \bigintsss_{z_1}^{z_2}\biggl(1-\dfrac{1}{1+ \biggl(\gamma_{j} {\beta_D } (\dfrac{{r_j}}{r})^{-\alpha}\biggl)^{}}\biggl) r_{j} dr_j\biggl),
\end{align}
which can be further simplified as

\begin{align}
\mathcal{L}_{I_{J}}(s)=\exp\biggl(-2\pi\rho_J \bigintsss_{z_1}^{z_2}\biggl(\dfrac{1}{1+ \biggl((\gamma_{j} {\beta_D })^{-1} (\dfrac{{r_j}}{r})^{\alpha}\biggl)^{}}\biggl) r_{j} dr_j\biggl),
\end{align}
where $z_1$ and $z_2$ constitute
the effective attacking area of the jammers, s.t., $z_1 < z_2$, or the area from where jammers can significantly affect the transmissions.
 To make the expression elegant let $u=\biggl({r_{j}}/{r({\gamma_{j} {\beta_D })^{{1}/{\alpha}}}}\biggl)^{2} $, $z_l=\biggl(\dfrac{z_1}{r(\gamma_{j} \beta_D )^{{1}/{\alpha}}}\biggl)^2$ and $z_u=\biggl(\dfrac{z_2}{r(\gamma_{j} \beta_D )^{1/\alpha}}\biggl)^2$,
 then above expression can be written as:
\begin{align}
\mathcal{L}_{I_{J}}(s)=\exp\biggl({-\pi\rho_J }{r^2 (\gamma_{j} \beta_D )^{{2}/{\alpha}} } \bigintsss_{z_l}^{z_u}\dfrac{1}{1+(u^{\alpha/2})}du\biggl).
\end{align}
Integral is computed via Gauss-hypergeometric approximation, given as:
\begin{equation}
\begin{split}
& \mathcal{L}_{I_{J}}(s)= \\
   & \exp\biggl(\dfrac{\pi \rho_J \gamma_{j}\beta_D r^{\alpha}}{(\alpha/2)-1} \biggl( z_2^{(2-\alpha)}{}_{2}F_{1}(1,1-{\dfrac{2}{\alpha}},2-{\dfrac{2}{\alpha}},-\gamma_j \beta_D (\dfrac{r}{z_2})^\alpha)\\
&    - z_1^{(2-\alpha)} {}_{2}F_{1}(1,1-{\dfrac{2}{\alpha}},2-{\dfrac{2}{\alpha}},-\gamma_j \beta_D (\dfrac{r}{z_1})^\alpha) \biggl)\biggl).
\end{split}
\end{equation}
